\documentclass[longauth]{aa} % for the long lists of affiliations 

\newcommand{\ttt}[1]{\times10^{#1}}

\usepackage[switch]{lineno}
\usepackage{graphicx}
\usepackage{txfonts}
\usepackage{color}
\usepackage{fixltx2e}
\usepackage[english]{babel}

\newcommand{\komm}[1]{\textcolor{black}{#1}} % changes are normal colour

\newcommand{\fermilat}{\textit{Fermi}-LAT} 

\begin{document}

   \title{Very-high-energy $\gamma$-ray observations of novae and dwarf novae with the MAGIC telescopes}

% authors 27.02.2015  Format AA
%
\author{
M.~L.~Ahnen\inst{1} \and
S.~Ansoldi\inst{2} \and
L.~A.~Antonelli\inst{3} \and
P.~Antoranz\inst{4} \and
A.~Babic\inst{5} \and
B.~Banerjee\inst{6} \and
P.~Bangale\inst{7} \and
U.~Barres de Almeida\inst{7,}\inst{26} \and
J.~A.~Barrio\inst{8} \and
J.~Becerra Gonz\'alez\inst{9,}\inst{27}
% ...
}
\institute { 
ETH Zurich, CH-8093 Zurich, Switzerland
\and Universit\`a di Udine, and INFN Trieste, I-33100 Udine, Italy
\and INAF National Institute for Astrophysics, I-00136 Rome, Italy
\and Universit\`a  di Siena, and INFN Pisa, I-53100 Siena, Italy
\and Croatian MAGIC Consortium, Rudjer Boskovic Institute, University of Rijeka and University of Split, HR-10000 Zagreb, Croatia
\and Saha Institute of Nuclear Physics, 1\textbackslash AF Bidhannagar, Salt Lake, Sector-1, Kolkata 700064, India
\and Max-Planck-Institut f\"ur Physik, D-80805 M\"unchen, Germany
\and Universidad Complutense, E-28040 Madrid, Spain
\and Inst. de Astrof\'isica de Canarias, E-38200 La Laguna, Tenerife, Spain
\and University of \L\'od\'z, PL-90236 Lodz, Poland
\and Deutsches Elektronen-Synchrotron (DESY), D-15738 Zeuthen, Germany
\and IFAE, Campus UAB, E-08193 Bellaterra, Spain
\and Universit\"at W\"urzburg, D-97074 W\"urzburg, Germany
\and Centro de Investigaciones Energ\'eticas, Medioambientales y Tecnol\'ogicas, E-28040 Madrid, Spain
\and Universit\`a di Padova and INFN, I-35131 Padova, Italy
\and Institute of Space Sciences, E-08193 Barcelona, Spain
\and Technische Universit\"at Dortmund, D-44221 Dortmund, Germany
\and Unitat de F\'isica de les Radiacions, Departament de F\'isica, and CERES-IEEC, Universitat Aut\`onoma de Barcelona, E-08193 Bellaterra, Spain
\and Universitat de Barcelona, ICC, IEEC-UB, E-08028 Barcelona, Spain
\and Japanese MAGIC Consortium, ICRR, The University of Tokyo, Department of Physics and Hakubi Center, Kyoto University, Tokai University, The University of Tokushima, KEK, Japan
\and Finnish MAGIC Consortium, Tuorla Observatory, University of Turku and Department of Physics, University of Oulu, Finland
\and Inst. for Nucl. Research and Nucl. Energy, BG-1784 Sofia, Bulgaria
\and Universit\`a di Pisa, and INFN Pisa, I-56126 Pisa, Italy
\and ICREA and Institute of Space Sciences, E-08193 Barcelona, Spain
\and Universit\`a dell'Insubria and INFN Milano Bicocca, Como, I-22100 Como, Italy
\and now at Centro Brasileiro de Pesquisas F\'isicas (CBPF\textbackslash MCTI), R. Dr. Xavier Sigaud, 150 - Urca, Rio de Janeiro - RJ, 22290-180, Brazil
\and now at NASA Goddard Space Flight Center, Greenbelt, MD 20771, USA and Department of Physics and Department of Astronomy, University of Maryland, College Park, MD 20742, USA
% ...
}

%   \date{Received September 15, 1996; accepted March 16, 1997}
   \date{}
  \abstract
   {
     In the last five years the \textit{Fermi} Large Area Telescope (LAT) instrument detected GeV $\gamma$-ray emission from five novae.
     The GeV emission can be interpreted in terms of an inverse Compton process of electrons accelerated in a shock.
     In this case it is expected that protons in the same conditions can be accelerated to much higher energies. 
     Consequently they may produce a second component in the $\gamma$-ray spectrum at TeV energies.
   }
   {
     We aim to explore the very-high-energy domain to search for $\gamma$-ray emission above 50\,GeV and to shed light on the acceleration process of leptons and hadrons in nova explosions. 
   }
   {
     We have performed observations with the MAGIC telescopes of the classical nova V339 Del shortly after the 2013 outburst, triggered by optical and subsequent GeV $\gamma$-ray detections. 
     \komm{We also briefly report on VHE observations} of the symbiotic nova YY Her and the dwarf nova ASASSN-13ax. 
     We complement the TeV MAGIC observations with the analysis of contemporaneous \fermilat\ data of the sources. 
     The TeV and GeV observations are compared in order to evaluate the acceleration parameters for leptons and hadrons. 
   }
   {
     No significant TeV emission was found from the studied sources. 
     We computed upper limits on the \komm{spectrum} and night-by-night flux. 
     The combined GeV and TeV observations of V339 Del \komm{limit} the ratio of \komm{proton to electron luminosities} to $L_p\lesssim 0.15 \, L_e$.
   }
   {}


   \maketitle
%
%________________________________________________________________

\section{Introduction}
A classical nova is a thermonuclear runaway leading to the explosive ejection of the envelope accreted onto a white dwarf (WD) in a binary system in which the companion is either filling or nearly filling its Roche surface \citep{be08,st12,wr14}.
They are a type of cataclysmic variable, i.e. optically variable binary systems with a mass transfer from a companion star to WD.
Novae are typically detected first in optical observations when the brightness of the object increases by 7-16 magnitudes. 
A thermal X-ray continuum is often seen in the energy spectra of novae.  
Symbiotic novae, like classical novae, are initiated by a thermonuclear explosion on the surface of the WD.
However in the case of symbiotic novae the WD is deeply immersed in the wind of a late-type companion star (see e.g. \citealp{sh11,sh12}).

The diffusive shock acceleration at the blast wave of symbiotic novae was expected to accelerate particles up to energies of a few TeV \citep{th07}.
In 2010 the first GeV $\gamma$-ray emission was detected by the \textit{Fermi} Large Area Telescope (LAT) from the symbiotic nova V407 Cyg \citep{ab10}.
The $\gamma$-ray emission can be explained in terms of leptonic or hadronic models \citep{ab10,novaescience}. 
The local radiation fields create a target for the inverse Compton (IC) scattering of the electrons.
Protons accelerated in the same conditions can interact with the matter producing $\gamma$-rays via proton-proton interactions.
Several models have been put forward to explain the GeV radiation. 
For instance, \cite{sb12} attribute the GeV $\gamma$-ray emission to the IC process on the strong radiation field of the red giant.
The same model predicts a second component in the TeV range due to proton-proton interactions with the wind of the red giant.
\komm{Alternatively,} \citet{md13} consider acceleration of leptons and hadrons in the nova shock.
In that model the magnetic field, which determines the acceleration efficiency, is obtained assuming an equipartition with the thermal energy density upstream of the shock resulting in the maximum energy of protons estimated to be $\sim 300$\, GeV.
The GeV $\gamma$-ray emission is then expected mostly from leptonic processes, namely the IC scattering of the nova light by the electrons accelerated in the shock.

In the last three years \fermilat\ has discovered GeV $\gamma$-ray emission from four additional novae, V1324 Sco, V959 Mon, V339 Del, and V1369 Cen \citep{cjs13,novaescience}.
In addition \fermilat\ has reported V745 Sco and Nova Sgr 2015 No. 2 as lower significance candidates \citep{cjs14,cc15}. 
Most of these sources are classical novae. 
Contrary to the symbiotic ones, they do not exhibit a strong wind of the companion star, but still they all show similar spectral properties. 
In classical novae particle acceleration can occur, for example, in a bow shock between the nova ejecta and the interstellar medium or in weaker internal shocks due to inhomogeneity of the nova ejecta \citep{novaescience}.
In particular, the orbital motion of the system can shape the nova ejecta into a faster polar wind and a slower region of denser material along the equatorial plane \citep{cho14}.
\cite{me15} suggest that the $\gamma$-ray emission might come from hadronic interactions of a faster outflow with an ejected shell.

So far no very-high-energy (VHE; $E>$100 GeV) $\gamma$-ray emission has been detected from any nova event. 
VERITAS observations of the symbiotic nova V407 Cyg beginning 10 days after the nova explosion yielded a differential upper limit on the flux at 1.6 TeV of $2.3 \times 10^{-12}\, \mathrm{erg\,cm^{-2}\,s^{-1}}$\citep{al12}.
 
Since late 2012 the MAGIC collaboration has been conducting a nova follow-up program in order to detect a possible VHE $\gamma$-ray component. 
At first the program focused on symbiotic novae. 
After the reports of the detection of GeV emission from classical novae by the \fermilat\, the program was extended also to bright classical novae and opened up to additional outbursts from other cataclysmic variables. 

In this paper we report on the observations performed with the MAGIC telescopes of V339 Del and present an updated analysis of \fermilat\ data contemporaneous with those observations. 
In Section~\ref{sec:ins} we describe the MAGIC telescopes and the \fermilat\ instrument.
The observations of V339 Del are presented in Section~\ref{sec:obs}.
In Section~\ref{sec:mod} we discuss the results of the GeV-TeV observations of V339 Del in terms of a hadronic-leptonic model. 
In addition to the classical nova V339 Del, MAGIC observed also dwarf nova ASASSN-13ax and symbiotic nova YY Her. 
Neither GeV nor TeV $\gamma$-ray emission was detected from those two sources.
The upper limits on these two objects are summarized in \komm{the online-only} Appendix \ref{appendix} while the \komm{main text} of the paper focuses on V339 Del.

\section{Instruments} \label{sec:ins}
V339 Del and the other outbursts observed by MAGIC were first detected and observed by optical instruments. 
The analysis of quasi-simultaneous \fermilat\ observations provides additional context for the MAGIC results. 

\subsection{MAGIC telescopes}
The VHE $\gamma$-ray observations were obtained using the MAGIC telescopes. 
MAGIC is a system of two 17\,m Cherenkov telescopes located on the Canary Island of La Palma at a height of 2200 m a.s.l. \komm{\citep{mup1}}.
The telescopes record $\gamma$-rays with energies above $\sim$50\,GeV. 
The sensitivity of the MAGIC telescopes in the best energy range ($\gtrsim300\,$GeV) is $\sim 0.6\%$ of Crab Nebula flux in 50\,h of observations \citep{mup2}.
The data were analyzed using the standard analysis chain \citep{magic_mars, mup2}.
The significance of a $\gamma$-ray excess was computed according to Eq.~17 from \citet{lm83}.
The upper limits on the flux were calculated following the approach of \cite{ro05} using 95\% confidence level (C.L.) and accounting for a possible 30\% systematic uncertainty on the effective area of the instrument \citep{mup2}. 

\subsection{\fermilat}
The \fermilat , a space-based, pair-conversion telescope, detects photons with energies from $20\,$MeV to $>300$\,GeV \citep{at09}.
We analyzed the LAT data in the energy range 100\,MeV $-$ 300\,GeV using an unbinned maximum likelihood method \citep{mat96} as implemented in the \textit{Fermi} Science Tools v9r32p5. We applied the P7REP\_SOURCE\_V15 LAT Instrument Response Functions (IRFs) and used the associated standard Galactic and isotropic diffuse emission models matched to the Pass 7 reprocessed Source class event selection\footnote{The P7REP data, IRFs, and diffuse models (gll\_iem\_v05\_rev1.fit and iso\_source\_v05.txt) are available at http://fermi.gsfc.nasa.gov/ssc.}.
We selected events within a region of interest (ROI) of $15^\circ$ in radius centered on the LAT best-fit position reported by \citet{novaescience} for V339 Del and required a maximum zenith angle of $100^\circ$ to avoid contamination by Earth limb photons. Because some of the LAT data were acquired during pointed mode observations, we applied an appropriate filter\footnote{http://fermi.gsfc.nasa.gov/ssc/data/analysis/documentation/Cicerone/\\Cicerone\_Likelihood/Exposure.html}, selecting good quality data at times when either the rocking angle was less than $52^\circ$ or the edge of the analysis region did not exceed the maximum zenith angle at $100^\circ$.
Sources from the 2FGL catalogue \citep{nol12} located within the ROI were included in the model used to perform the fitting procedure.

\section{Observations of V339 Del and results}
\label{sec:obs}
\begin{figure}
\begin{minipage}{0.49\textwidth}
\includegraphics[width=0.99\textwidth]{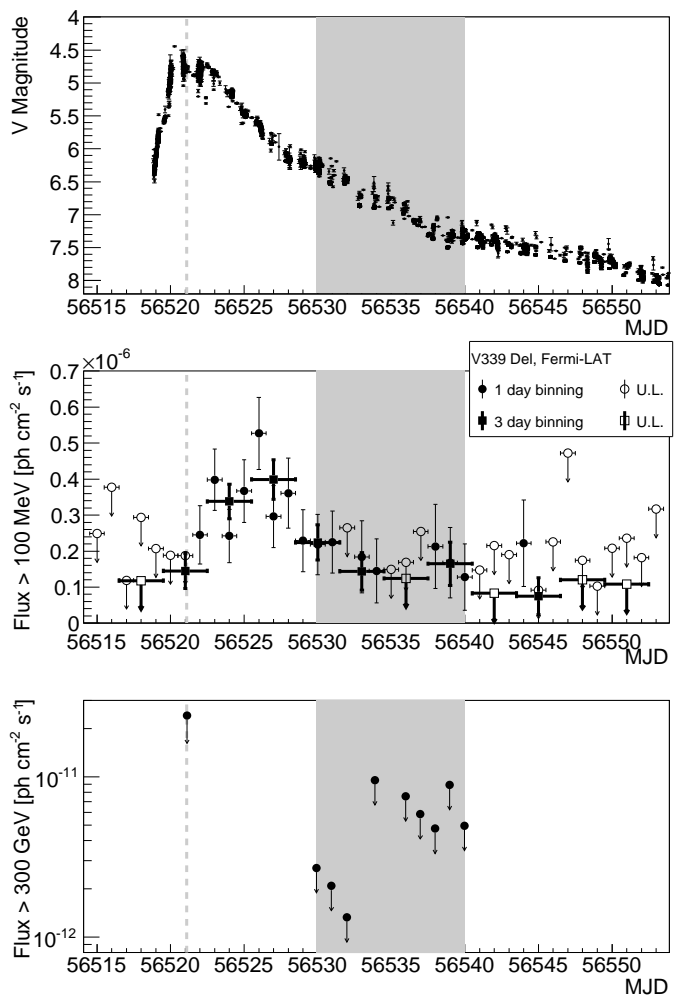}
\caption{
Multi-wavelength light curve of V339 Del during the outburst in August 2013.
Top panel: Optical observations in the V band obtained from the AAVSO-LCG\protect\footnote{http://www.aavso.org/lcg} service.
Middle panel: The \fermilat\ flux (filled symbols) and upper limits (empty symbols) above 100 MeV in 1-day (circles, thin lines) or 3-day (squares, thick lines) bins.
A 95\% C.L. flux upper limit is shown for time bins with TS$<$4.
Bottom panel: Upper limit on the flux above 300 GeV observed with MAGIC telescopes.
The gray band shows the observation nights with MAGIC. 
The dashed gray line shows a MAGIC observation night affected by bad weather.}
\label{fig:del_mwl}
\end{minipage}
\end{figure}

V339 Del was a fast, classical CO nova detected by optical observations on 2013 August 16 (CBET \#3628), MJD 56520. 
The nova was exceptionally bright reaching a magnitude of V$\sim 5\,$mag (see top panel of Fig.~\ref{fig:del_mwl}), and it triggered follow-up observations at frequencies ranging from radio to VHE $\gamma$-rays.
Photometric measurements suggest a distance for V339 Del of $4.5\pm0.6$\,kpc \citep{sch14}.
Nearly a month after the optical detection, X-ray emission was detected in the 1--10 keV energy band by {\it Swift}/XRT \citep{pa13}.
Afterwards, the object became a low energy X-ray source with most of the photons detected in the 0.3--1 keV energy range \citep{os13}.
The object shows large amplitude variations and a 54~s quasi-periodic oscillation in the soft X-ray energy band.
These are possibly explained by the spin of the white dwarf or an oscillation in the nuclear burning rate \citep{be13,ne13}.
The spectroscopic observations performed on 2013 August 18 revealed emission wings extending to about $\pm 2000\,$km/s and a Balmer absorption component at a velocity of $(600\pm 50)\,$km/s \citep{sh13}.
The pre-outburst optical images revealed the progenitor of nova V339 Del to be a blue star \citep{de13}.

Originally, MAGIC observations of V339 Del were motivated by its extreme optical outburst.
The subsequent detection of GeV emission by the \fermilat\ from the direction of V339 Del added incentive for VHE observations.
MAGIC acquired data starting on the night of 2013 August 16 but these were marred by poor weather. 
The good quality data used for most of the analysis spans 8 nights between 2013 August 25 and September 4. 
The total effective time was 11.6\,h.
In addition to the nightly upper limits we performed a dedicated analysis of the poor-quality (affected by Calima, a dust layer originating from Sahara) night of 2013 August 16.
We applied an estimated energy and collection area correction based on LIDAR measurements \citep{fr14}. 
No VHE $\gamma$-ray signal was found from the direction of V339 Del.
We computed night-by-night integral upper limits above 300\,GeV (see bottom panel of Fig.~\ref{fig:del_mwl}) and differential upper limits for the whole good quality data set in bins of energy (see Section \ref{sec:model}). 

Nova V339 Del was the subject of a \textit{Fermi} Target of Opportunity (ToO) observation \citep{AT5302} triggered by the optical discovery (CBET \#3628); a pointed observation favoring the nova started on 2013 August 16 and lasted for 6 days.
The $\gamma$-ray emission from V339 Del was first detected by \fermilat\ in a 1-day bin on 2013 August 18 \citep{novaescience}. 
The emission peaked on 2013 August 22 and entered a slow decay phase afterwards (Fig.~\ref{fig:del_mwl}).
We fitted the flux for the light curves shown in the middle panel of Fig.~\ref{fig:del_mwl} by assuming a power-law spectral model with the normalization left free to vary and the photon index fixed to a single value.
We selected a fixed value for the photon index of 2.3 by calculating the average of the most significant 1-day bins (Test Statistic values TS$>$9)\footnote{The source significance (in sigmas) is $\sim$ $\sqrt{TS}$ assuming one degree of freedom}.  

\begin{figure*}
\centering
\includegraphics[width=0.69\textwidth]{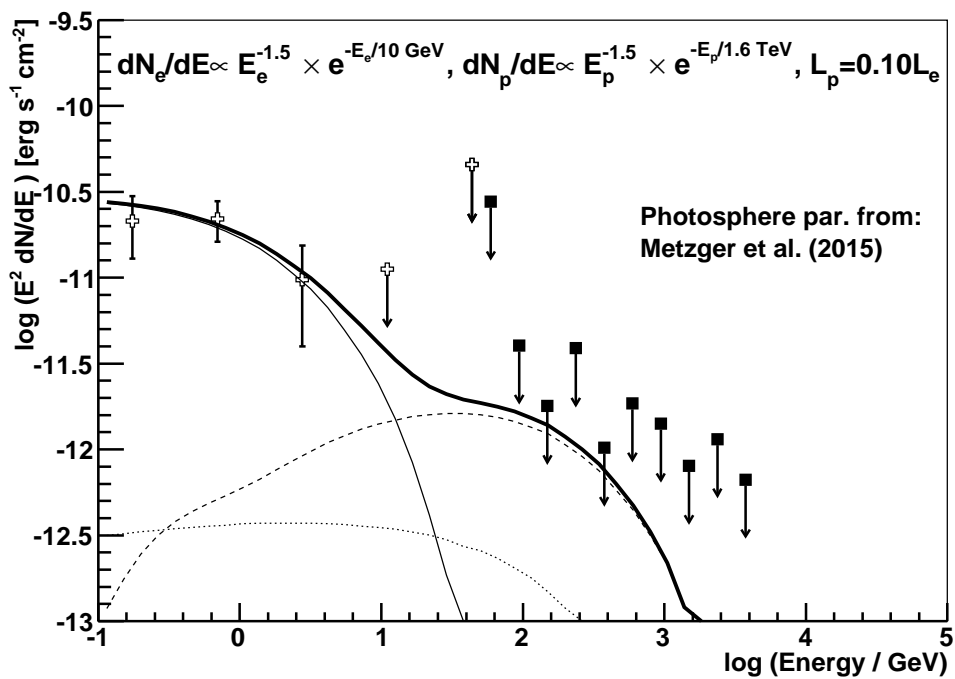}
\caption{
Differential upper limits on the flux from V339 Del as measured by MAGIC (filled squares) and the flux measured by \fermilat\ (empty crosses) in the same time period, 2013 August 25 to September 4.
The thin solid line shows the IC scattering of thermal photons in the nova's photosphere.
The dashed line shows the $\gamma$-rays coming from the decay of $\pi^0$ from hadronic interactions of the relativistic protons with the nova ejecta.
The dotted line shows the contribution of $\gamma$-rays coming from IC of e$^+$e$^-$ originating from $\pi^+\pi^-$ decays.
Thick solid lines show the total predicted spectrum. 
The total energy of electrons is $6\times 10^{41}$\,erg and the assumed proton to electron luminosity ratio is $L_p/L_e=0.1$. 
Electrons and protons are injected with a power law with a spectral index of $1.5$ and the cut-offs reported in the figures. 
Photosphere parameters (see Table \ref{tab:par}) obtained from \citet{me15}.
}\label{fig:del2013_sed}
\end{figure*}
The spectral energy distribution (SED) for V339 Del shown in Fig.~\ref{fig:del2013_sed} was extracted in 5 logarithmically spaced energy bins from 100 MeV to 100 GeV.
Similarly to the light curves, energy-binned data shown in Fig.~\ref{fig:del2013_sed} were fitted using a power law and calculating a 95\% C.L. upper limit for bins with TS$<$9. 
In the period coincident with the MAGIC observations (2013 August 25 to September 4), the \fermilat\ spectrum can be described by a power law with an index of $2.4\pm0.2$ and flux above 100 MeV of $(0.15\pm 0.04) \times 10^{-6} \mathrm{\,ph\,cm^{-2}\,s^{-1}}$. 
Only statistical error is included since that dominated over systematic error in these data.
The spectral fit for this period had a TS of 49 and did not permit a constraint on an exponential cut-off in energy.
The \fermilat\ analysis for the full decay phase, 2013 August 22 to September 12 (MJD 56526-56547), provided a more significant signal with a TS of 121 and a similar value of flux above 100 MeV, $(0.13\pm 0.03) \times 10^{-6} \mathrm{\,ph\,cm^{-2}\,s^{-1}}$. 
The spectrum for the longer time period was fitted by an exponentially cut-off power law with an index of $1.4\pm 0.3$ and a cut-off energy of $1.6\pm0.8$\,GeV. 
The fit improved in significance by $3.3\sigma$ with respect to a power-law model. 
The most energetic photon associated with V339 Del (89\% probability for the best-fit model) had an energy $E=5.9$\,GeV and was recorded on 2013 August 30, i.e., within the time period covered by the MAGIC observations.

\section{Modeling of $\gamma$-rays from nova V339 Del}\label{sec:mod}
\label{sec:model}
Of the 3 objects observed by the MAGIC telescopes and discussed in this paper, V339 Del is the only one detected at GeV energies by the \fermilat.
\komm{Extensive optical observations provided constraints on the} companion star and the photosphere of the nova.
Therefore it has the highest potential for constraining the leptonic and hadronic processes in novae, and we concentrate the modeling efforts on it. 
We follow the modified model of \cite{sb12}.
The original scenario was applied to the symbiotic nova V407 Cyg. 
In that case the GeV $\gamma$-ray emission was attributed to IC of the electrons on the strong radiation field in the vicinity of the red giant companion star.
For V339 Del, however, the radiation field of the companion star is not as strong. 
The photosphere of the nova provides a dominant target for the IC process. 
Moreover, the wind of the companion star is not as dense as in the V407 Cyg symbiotic system.
Nevertheless, if protons of sufficient energy are accelerated in the nova shock, they can interact with the ejecta of the nova producing pions that decay to $\gamma$-rays. 
As in \cite{sb12} the GeV $\gamma$-ray emission can be used to constrain the parameters describing the acceleration of the electrons, which are otherwise poorly known. 
Protons will then be accelerated in the same conditions but up to much higher energies due to lower energy losses. 

In order to apply the above model we need to evaluate the radiation field which is encountered by electrons at the time of the observations by MAGIC and \fermilat .
First, we need to estimate the parameters of the nova photosphere about 10 days after the optical detection. 
The spectral evolution of the V339 Del was studied by \cite{me15}.
During the time of the observations by the MAGIC telescope the reported temperature of the photosphere was $\sim7000$\,K.
In the early phases of the nova (fireball stage) the optical observations were consistent with a pseudophotosphere with a temperature of about 10000\,K.
With a reported optical luminosity of $\sim 6 \times 10^{4}L_\odot$ this corresponds to a radius of the photosphere of $1.2\times 10^{13}$\,cm (see Table~\ref{tab:par}).
Following the approach of \citet{sh13b} we also used a second set of photosphere parameters, in which UV emission, not available at this late stage of V339 outburst is estimated using measurements of a similar source, namely OS And.

Let us consider a photosphere of a radius $R_{ph} = 10^{13} R_{ph, 13}\,$cm and a temperature $T_{ph} = 10^{4} T_4\,$K.
The leptons and hadrons are accelerated at a distance of $R_{sh}=10^{13} R_{sh,13}$\,cm from the photosphere.
Assuming that the velocity of the shock is $\sim1000\,\mathrm{km/s}$, similar to the one observed in another GeV-emitting nova \citep{cho14}, we estimate that during the MAGIC observations the distance from the WD to the shock was $R_{sh}\sim 10^{14}\mathrm{cm}$.
The $\gamma$-rays with energy $E_\gamma = 10 E_{10}$\,GeV can be produced via IC scattering in the Thomson regime of thermal photons by electrons with energy of
\begin{equation}
  E_e = 22\times E_{10}^{1/2} / \left( T_4 (1+\cos\beta)\right)^{1/2} \, \mathrm{[GeV]},
\end{equation} 
where $\beta$ is the angle between the electron and the direction to the point on the photosphere where the thermal photon was emitted.
On the other hand, by comparing the energy losses from the IC scattering with the acceleration rate we obtain a maximum energy for electrons of (see e.g. \citealp{sb12})
\begin{equation}\label{eq:ee}
  E_{e,\max} = 13 (\xi_{-4} B)^{1/2} R_{sh} / (T_4^2 R_{ph}) \, \mathrm{[GeV]},
\end{equation}
where $B$ is the magnetic field at the shock (measured in Gauss), and $\xi = 10^{-4} \xi_{-4}$ is the acceleration coefficient. 
The acceleration coefficient is defined by the acceleration time $\tau_{acc}=1\,E/\xi_{-4} B$.
By comparing the above two formulae we obtain
\begin{equation}
  \xi_{-4} B = 2.9 E_{10} T_4^3 R_{ph}^2 / (R_{sh}^2 (1+\cos\beta)).
\end{equation}

In the same conditions protons with energy $E_p$  (measured in units of GeV) can be accelerated with a time scale of
\begin{equation}
  \tau_{acc,p}=1 E_p /(\xi_{-4} B) = 
  0.34 \frac{E_p R_{sh}^2 (1+\cos\beta)}{E_{10}  T_4^3 R_{ph}^2}  \, \mathrm{[s]}.
\end{equation}
The acceleration can be limited by the dynamic time scale of $t_d\sim 10\,$days (after which time most of the MAGIC observations were performed) or by the energy losses of protons from the \textit{pp} collisions. 
The time scale of the latter can be computed as
\begin{equation}
  \tau_{pp} = (\sigma_{pp} n_H k c)^{-1},
\end{equation}
where $\sigma_{pp}\approx 3\times 10^{-26}\mathrm{cm^2}$ is the interaction cross section, $n_H$ is the density of the nova ejecta, the inelasticity coefficient, $k\approx0.5$, is the fraction of energy lost in each interaction, and $c$ is the speed of light.
The density of the ejecta will decrease as the nova shock progresses with the speed of $v=10^{3}v_3\,\mathrm{km\, s^{-1}}$ following
\begin{equation}
  n_H = 4.4\times 10^{12}  M_{-5} / (v_3^3 t_d^3)\ \mathrm{[cm^{-3}]}, %% 4.41e12
\end{equation}
where $10^{-5} M_{-5} M_\odot$ is the total mass ejected during the outburst.
Thus,
\begin{equation}
  \tau_{pp} =  500 (v_3^3 t_d^3)/M_{-5} = 780 R_{sh,13}^3/M_{-5}  \, \mathrm{[s]}, %%505, 783
\end{equation}
where we used that $R_{sh}=v (86400\,t_{d})$\ \komm{$\approx 8.6 \times 10^{13}\mathrm{cm}$}.

By comparing the acceleration and cooling time scales we obtain
\begin{equation}\label{eq:ep}
  E_p = 2300 \frac{E_{10} T_4^3 R_{ph,13}^2 R_{sh,13}}{(1+\cos\beta)M_{-5}}  \, \mathrm{[GeV]}. %%2290
\end{equation}

\begin{table}[!t]
\begin{center}
\caption{
Parameters characterizing the optical emission of V339 Del (photosphere temperature $T_4 \times 10^4\,$K, radius $R_{ph,13} \times 10^{13}\,$cm and luminosity $L$) according to the two scenarios assumed in the modeling of the GeV and TeV emission.}
\label{tab:par}
\begin{tabular}{c|c c c}
     & $T_4$ & $R_{ph,13}$ & $L/L_\odot$ \\ \hline
\cite{me15} & 0.7 & 1.2  & $6 \times 10^{4}$\\
optical+UV  & 1.3 & 0.4  & $8 \times 10^{4}$ \\
\end{tabular}

\end{center}
\end {table}

Using the parameters of the photosphere obtained by \cite{me15} (see Table~\ref{tab:par}) a fortnight after the peak and taking into account the break in the GeV spectrum at $\sim 1.6\,$GeV, it is plausible to expect protons to be accelerated at least up to energies of $\sim 1.6\,$TeV.
The maximum energy of the protons is expected to be lower, 1.1\,TeV, in the case of the second set of photosphere parameters computed from the combined optical and UV fit.

The acceleration can be also limited by the dynamic time scale instead of the hadronic interaction losses. 
By comparing the dynamic time scale with the proton cooling time scale we find that the former starts to dominate only after $13 M_{-5}^{1/2} v_3^{-3/2}$ days. 
Therefore the accelerated protons will mostly cool down on the time scale of the MAGIC observations due to energy losses in hadronic interactions.  
The normalization of both components is determined by $L_p/L_e$, i.e. the ratio of the total power of accelerated protons to that of electrons. 

We used the numerical code of \cite{sb12} to model the \fermilat\ GeV spectrum and to compare the sub-TeV predictions with the MAGIC observations.
We consider that the electrons and protons accelerated in Fermi-like acceleration obtain a power-law energy spectrum with a spectral index of $1.5$.
The spectra of electrons and protons cut off at energies determined by equations \ref{eq:ee} and \ref{eq:ep} respectively.
In Fig.~\ref{fig:del2013_sed} we show the predictions for leptonic or hadronic spectra compared with the \fermilat\ and MAGIC measurements.

The \fermilat\ spectrum can be described mostly by IC scattering of the thermal photons in the nova's photosphere by electrons. 
The expected hadronic component overpredicts the MAGIC observations at $\sim 100 \,$GeV by a factor of a few for the case of equal power of accelerated protons and electrons (i.e. $L_p=L_e$).
Using the upper limits from the MAGIC observations we can place the limit on $L_p\lesssim 0.15 L_e$.
We checked that our results are unchanged when using radiation field parameters resulting in a significantly different emission in the UV range, which is not constrained by the observations of the V339 nova at the time of the MAGIC observations.

\komm{The increased power in electrons compared to protons may be related to how particles with different mass are injected in the acceleration process.}
Interestingly, the appearance of energetic $\mathrm{e^+e^-}$ pairs from nuclear decays produced in the nova explosion could help to inject them preferentially into the shock acceleration
process.
On the other hand, \citet{sch03} suggests that in a low-beta plasma acceleration of electrons is preferred over protons if the particles are accelerated out of a thermal population.
Both effects could lower the $L_p/L_e$ ratio. 

\section{Conclusions}\label{sec:conc}
No VHE $\gamma$-ray emission was found from the direction of V339 Del.
The contemporaneous \fermilat\ observations revealed GeV emission from V339 Del.
We modeled this GeV emission as the IC of thermal photons from the photosphere by the GeV electrons accelerated in the nova shock.
We used the \fermilat\ and MAGIC observations of V339 Del to constrain the number of protons accelerated in the same conditions as the electrons in the nova shock.
The modeling shows that the total power of accelerated protons must be $\lesssim 15\%$ of the total power of accelerated electrons.
MAGIC will continue to observe promising $\gamma$-ray nova candidates in the following years.

\begin{acknowledgements}
The MAGIC Collaboration would like to thank
the Instituto de Astrof\'{\i}sica de Canarias
for the excellent working conditions
at the Observatorio del Roque de los Muchachos in La Palma.
The financial support of the German BMBF and MPG,
the Italian INFN and INAF,
the Swiss National Fund SNF,
the ERDF under the Spanish MINECO, and
the Japanese JSPS and MEXT
is gratefully acknowledged.
This work was also supported
by the Centro de Excelencia Severo Ochoa SEV-2012-0234, CPAN CSD2007-00042, and MultiDark CSD2009-00064 projects of the Spanish Consolider-Ingenio 2010 programme,
by grant 268740 of the Academy of Finland,
by the Croatian Science Foundation (HrZZ) Project 09/176 and the University of Rijeka Project 13.12.1.3.02,
by the DFG Collaborative Research Centers SFB823/C4 and SFB876/C3,
and by the Polish MNiSzW grant 745/N-HESS-MAGIC/2010/0.
and NCN 2011/01/B/ST9/00411. %% WB & JS
JS is supported by Fundacja U\L .
We thank M.A. P\'erez-Torres for the information about nova ASASSN-13ax.

The \textit{Fermi} LAT Collaboration acknowledges generous ongoing support
from a number of agencies and institutes that have supported both the
development and the operation of the LAT as well as scientific data analysis.
These include the National Aeronautics and Space Administration and the
Department of Energy in the United States, the Commissariat \`a l'Energie Atomique
and the Centre National de la Recherche Scientifique / Institut National de Physique
Nucl\'eaire et de Physique des Particules in France, the Agenzia Spaziale Italiana
and the Istituto Nazionale di Fisica Nucleare in Italy, the Ministry of Education,
Culture, Sports, Science and Technology (MEXT), High Energy Accelerator Research
Organization (KEK) and Japan Aerospace Exploration Agency (JAXA) in Japan, and
the K.~A.~Wallenberg Foundation, the Swedish Research Council and the
Swedish National Space Board in Sweden.
Additional support for science analysis during the operations phase is gratefully acknowledged
from the Istituto Nazionale di Astrofisica in Italy and the Centre National d'\'Etudes Spatiales in France.

Authors would like to thank S. N. Shore for scientific discussions and providing an optical spectrum of V339 Del and UV spectrum of OS And. 
%We would also like to thanks the anonymous referee for the comments which helped improving the paper.
We would also like to thank the anonymous referee for comments that helped to improve the paper.
\end{acknowledgements}

%-------------------------------------------------------------------

\clearpage 
\appendix
\section{\komm{(Online only)} YY Her and ASASSN-13ax}\label{appendix}
Two additional objects, the symbiotic nova YY Her and dwarf nova ASASSN-13ax, were observed as part of the broader campaign but neither was detected to have GeV or TeV emission. 
In this appendix we report the values of the differential upper limits from those two sources. 
The \fermilat\ analysis was configured as described for V339 Del with the ROIs centered on the optical positions for YY Her and ASASSN-13ax. 
Data were selected within the time windows reported, preceding the optical peak and containing approximately the following two weeks of observations. 
The region model was derived from the 2FGL catalogue and contains all sources within $20^\circ$ of the ROI center. 
The spectra for YY Her and ASASSN-13ax were described using a power law with the photon index fixed to 2.2.
The flux normalization for the source of interest and the normalization for the isotropic template were left free to vary while the other source parameters were fixed, either to catalogue values or refitted values if the source flux was detected in the analysis window and deviated by more than $1\sigma$ from the catalogue value. 
The upper limits were calculated at 95\% C.L. using the Bayesian method provided with the \textit{Fermi} Science Tools\footnote{http://fermi.gsfc.nasa.gov/ssc/data/analysis/scitools/python\_tutorial.html}.

The upper limits from MAGIC observations were computed in 5 bins per decade in energy. 
The photon spectral index assumed for both sources was 2.6.

YY Her is a symbiotic nova system that undergoes a recurrent pattern of outbursts. 
MAGIC observations of YY Her occurred on the night of April 22, 7 days after the optical maximum. 
The observations were motivated by the symbiotic nature of the source.
In the absence of detectable emission, flux upper limits at 95\% C.L. were calculated to be $5.0\times \mathrm{10^{-12} \,ph\,cm^{-2}\,s^{-1}}$ above 300 GeV. 
Emission was not detected in the LAT over the interval 2013 April 10 to April 30 (MJD 56392.5 to 56412.5). 
Upper limits at 95\% C.L. were calculated to be $2.8 \times 10^{-8} \mathrm{\,ph\,cm^{-2}\,s^{-1}}$ above 100\,MeV.
Differential upper limits obtained from the \fermilat\ and MAGIC observations of YY Her are summarized in Tab.~\ref{tab:yyher_ul}.

ASASSN-13ax is a member of a different class of cataclysmic variable, the dwarf novae, which are known for significantly weaker optical outbursts (typically 2-8 magnitudes) than classical novae. 
The source was observed by MAGIC due to its very strong optical outburst of 7.7 magnitudes \citep{st13}.
Instead of undergoing a thermonuclear explosion on the surface of the WD, these outbursts are caused by the gravitational energy release from an instability in the accretion disk surrounding the WD.
The MAGIC observations were performed on two consecutive nights starting on 2013 July 4, soon after the optical outburst seen on 2013 July 1.
In the absence of detectable VHE emission, upper limits at 95\% C.L. were calculated to be $1.5\times \mathrm{10^{-12} \,ph\,cm^{-2}\,s^{-1}}$ above 300 GeV.
Emission was not detected in the LAT over the interval 2013 June 25 to July 15 (MJD 56468.5 to 56488.5). 
Upper limits at 95\% C.L. were calculated to be $1.6 \times 10^{-8} \mathrm{\,ph\,cm^{-2}\,s^{-1}}$ above 100\,MeV. 
Differential upper limits obtained from the \fermilat\ and MAGIC observations of ASASSN-13ax are summarized in Tab.~\ref{tab:asassn_ul}

\begin{table}[t]
\begin{center}
\caption{Differential upper limits on the flux from YY Her as measured by the \fermilat\ and MAGIC.
The bins extend from $E_{min}$ to $E_{max}$ while the upper limit value is computed at the energy of $E_{UL}$.}
\label{tab:yyher_ul}

\begin{tabular}{c c c c }

\hline \hline
  \multicolumn{4}{c}{ \fermilat , MJD: 56392.5-56412.5}\\
  \hline
$E_{min}$ [GeV] & $E_{max}$ [GeV] & $E_{UL}$ [GeV] & $F_{UL}$ $[$TeV cm$^{-2}$s$^{-1}]$  \\
\hline
0.100 & 0.316 & 0.178 & 6.1$\ttt{-12}$ \\
0.316 & 1.00 & 0.562 & 4.2 $\ttt{-12}$\\
1.00 & 3.16 & 1.78 & 4.7 $\ttt{-12}$\\
3.16 & 10.0 & 5.62 & 1.2$\ttt{-11}$\\
10.0 & 100.0 & 31.6 & 1.5$\ttt{-11}$\\
\hline    
\hline
  \multicolumn{4}{c}{MAGIC, MJD=56405}\\
  \hline
50.0 & 79.2 & 59.4 & 7.1$\ttt{-11}$ \\
79.2 & 125.6 & 94.2 & 1.9 $\ttt{-11}$\\
125.6 & 199.1 & 149.3 & 3.8 $\ttt{-12}$\\
199.1 & 315.4 & 236.6 & 6.5$\ttt{-12}$\\
315.4 & 500.0 & 375.1 & 4.3$\ttt{-12}$\\

\hline
\end{tabular}
\end{center}
\end {table}

\begin{table}[t]
\begin{center}
\caption{Differential upper limits on the flux from ASASSN-13ax as measured by the \fermilat\ and MAGIC (see text for details).
Columns as in Table~\ref{tab:yyher_ul}.}
\label{tab:asassn_ul}

\begin{tabular}{c c c c }

\hline \hline
  \multicolumn{4}{c}{ \fermilat\ , MJD 56468.5 - 56488.5}\\
  \hline
$E_{min}$ [GeV] & $E_{max}$ [GeV] & $E_{UL}$ [GeV] & $F_{UL}$ $[$TeV cm$^{-2}$s$^{-1}]$  \\
\hline
0.100 & 0.316 & 0.178 & 3.3$\ttt{-12}$ \\
0.316 & 1.00 & 0.562 & 3.7 $\ttt{-12}$\\
1.00 & 3.16 & 1.78 & 2.6 $\ttt{-12}$\\
3.16 & 10.0 & 5.62 & 1.4$\ttt{-11}$\\
10.0 & 100.0 & 31.6 & 1.5$\ttt{-11}$\\
\hline    
\hline
  \multicolumn{4}{c}{ MAGIC, MJD: 56478-56479 }\\
  \hline
79.2 & 125.6 & 94.2 & 1.2 $\ttt{-11}$\\
125.6 & 199.1 & 149.3 & 1.2 $\ttt{-12}$\\
199.1 & 315.4 & 236.6 & 1.7$\ttt{-12}$\\
315.4 & 500.0 & 375.1 & 7.3$\ttt{-13}$\\
500.0 & 792.4 & 594.4 & 1.1$\ttt{-12}$\\
792.4 & 1255.9 & 942.1 & 7.9$\ttt{-13}$\\
1255.9 & 1990.5 & 1493.1 & 9.1$\ttt{-13}$\\
1990.5 & 3154.7 & 2366.4 & 1.7$\ttt{-12}$\\
3154.7 & 5000.0 & 3750.5 & 9.3$\ttt{-13}$\\

\hline
\end{tabular}
\end{center}
\end {table}


\begin{thebibliography}{}
\bibitem[Abdo et al.(2010)]{ab10} Abdo, A.~A., Ackermann, M., Ajello, M., et al.\ 2010, Science, 329, 817  % V407 Cygni Fermi
\bibitem[Ackermann et al.(2014)]{novaescience} Ackermann, M., Ajello, M., Albert, A., et al.\ 2014, Science, 345, 554-558  % Fermi Establishes Classical Novae ...
%\bibitem[Aleksi\'{c} et al.(2015a)]{mup1} Aleksi\'{c}, J., Ansoldi, S., Antonelli, L. A., et al.\ 2015a, accepted for publication in Astropart. Phys., DOI: 10.1016/j.astropartphys.2015.04.004, arXiv:1409.6073 % upgrade paper part I
%\bibitem[Aleksi\'{c} et al.(2015b)]{mup2} Aleksi\'{c}, J., Ansoldi, S., Antonelli, L. A., et al.\ 2015b, accepted for publication in Astropart. Phys., DOI: 10.1016/j.astropartphys.2015.02.005, arXiv:1409.5594 % upgrade paper part II 
\bibitem[Aleksi\'{c} et al.(2016a)]{mup1} Aleksi\'{c}, J., Ansoldi, S., Antonelli, L. A., et al.\ 2016a, Astroparticle Physics, 72, 61–75 % upgrade paper part I
\bibitem[Aleksi\'{c} et al.(2016b)]{mup2} Aleksi\'{c}, J., Ansoldi, S., Antonelli, L. A., et al.\ 2016b, Astroparticle Physics, 72, 76–94 % upgrade paper part II 
\bibitem[Aliu et al.(2012)]{al12} Aliu, E., Archambault, S., Arlen, T., et al.\ 2012, \apj, 754, 77 %% VERITAS nova cygni
\bibitem[Atwood et al.(2009)]{at09} Atwood, W.~B., Abdo, A. A., Ackermann, M., et al.\ 2009, \apj, 697, 1071  
\bibitem[Bode \& Evans(2008)]{be08} Bode, M.~F., \& Evans, A.\ 2008, Classical Novae, 2nd Edition.~Edited by M.F.~Bode and A.~Evans.~Cambridge Astrophysics Series, No.~43, Cambridge: Cambridge University Press, 2008.
\bibitem[Beardmore et al (2013)] {be13} Beardmore, A.~P., {Osborne}, J.~P., {Page}, K.~L \ 2013, The Astronomer's Telegram \# 5573 
\bibitem[Castro-Tirado et al.(2013)]{ct13} Castro-Tirado, A.~J., Martin-Carrillo, A., \& Hanlon, L.\ 2013, The Astronomer's Telegram, \#~5314 
\bibitem[Cheung et al.(2013b)]{cjs13} Cheung, C.~C., Jean, P., \& Shore, S.~N.\ 2013, The Astronomer's Telegram, \#~5653 %% V1369 Centauri 2013
\bibitem[Cheung et al.(2014a)]{cjs14} Cheung, C.~C., Jean, P., \& Shore, S.~N.\ 2014, The Astronomer's Telegram, \#~5879 %% V745 Sco
\bibitem[Cheung et al.(2015)]{cc15} Cheung, C.~C., Jean, P., \& Shore, S.~N., 2015, The Astronomer's Telegram, \#~7283 %% Nova Sgr 2015 No. 2
\bibitem[Chomiuk et al.(2014)]{cho14} Chomiuk, L., Linford, J.~D., Yang, J., et al.\ 2014, \nat, 514, 339
\bibitem[Copperwheat et al.(2013)]{co13} Copperwheat, C.~M., Bersier, D. F., Shappee, B. J., et al.\ 2013, The Astronomer's Telegram, 
\# 5195 %% optical spectrum of ASASSN-13ax
\bibitem[Deacon et al.(2014)]{de14} Deacon, N.~R., Hoard, D.~W., Magnier, E.~A., et al.\ 2014, \aap, 563, A129 % Del 2013 progenitor optical spectrum
\bibitem[Denisenko et al.(2013)]{de13} Denisenko, D., Jacques, C., Pimentel, E., et al.\ 2013, \iaucirc, 9258, 2 % Del 2013 progenitor optical
\bibitem[Fruck et al.(2014)]{fr14} Fruck, C., Gaug, M., Zanin, R., et al.\ 2014, Proc of 33rd ICRC, Rio de Janeiro, Brazil, arXiv:1403.3591 % LIDAR Correction
\bibitem[Hays et al.(2013)]{AT5302} Hays, E., {Cheung}, T., {Ciprini}, S., {The Fermi LAT Collaboration} \ 2013, The Astronomer's Telegram \# 5302
\bibitem[Herbig(1950)]{he50} Herbig, G.~H.\ 1950, \pasp, 62, 211 %% YY Her optical clasification
\bibitem[Li \& Ma(1983)]{lm83} Li, T.-P., \& Ma, Y.-Q. 1983, \apj, 272, 317
\bibitem[Martin \& Dubus(2013)]{md13} Martin, P., \& Dubus, G.\ 2013, \aap, 551, A37 %% nova model
\bibitem[Metzger et al.(2015)]{me15} Metzger, B.~D., Finzell, T., Vurm, I., et al.\ 2015, \mnras, 450, 2739
\bibitem[Mattox et al.(1996)]{mat96} Mattox, J.~R., Bertsch, D. L., Chiang, J., et al.\ 1996, \apj, 461, 396 %% Likelihood
\bibitem[Munari et al.(1997)]{mu97} Munari, U., Rejkuba, M., Hazen, M., et al.\ 1997, \aap, 323, 113 %% YY Her optical outbursts
\bibitem[Munari et al.(2013)]{mu13} Munari, U., Dallaporta, S., Castellani, F., et al.\ 2013, The Astronomer's Telegram \# 4996
\bibitem[Ness et al.(2013)]{ne13} Ness, J.~U., Schwarz, G. J., Page, K. L., et al. \ 2013, The Astronomer's Telegram \# 5626
\bibitem[Nolan et al.(2012)]{nol12} Nolan, P.~L., Abdo, A. A., Ackermann, M., et al. \ 2012, Astrophys. J. Supp. 199, 31 % 2nd LAT catalog
\bibitem[Osborne et al.(2013)]{os13} Osborne, J.~P., {Page}, K., {Beardmore}, A. et al. \ 2013, The Astronomer's Telegram \# 5505 %% , {Woodward}, C.
\bibitem[Page et al.(2013)]{pa13}  {Page}, K.~L. and {Beardmore}, A.~P. \ 2013, The Astronomer's Telegram \# 5429 
\bibitem[Rolke et al.(2005)]{ro05} Rolke, W.~A., L{\'o}pez, A.~M., \& Conrad, J.\ 2005, NIM A, 551, 493 
\bibitem[Schaefer et al.(2014)]{sch14} Schaefer, G.~H., Brummelaar, T.~T., Gies, D.~R., et al.\ 2014, \nat, 515, 234
\bibitem[Schlickeiser(2003)]{sch03} Schlickeiser, R.\ 2003, Energy Conversion and Particle Acceleration in the Solar Corona, 612, 230 % acceleration
\bibitem[Shore et al.(2011)]{sh11} Shore, S.~N., Wahlgren, G.~M., Augusteijn, T., et al.\ 2011, \aap, 527, AA98 % (Liimets, T., Page, K. L., Osborne, J. P., Beardmore, A. P., Koubsky, P., Slechta, M., Votruba, V.,) V407 Cyg
\bibitem[Shore et al.(2012)]{sh12} Shore, S.~N., Wahlgren, G.~M., Augusteijn, T., et al.\ 2012, \aap, 540, AA55  % ( Liimets, T., Koubsky, P., Slechta, M., Votruba, V.,)V407 Cyg
\bibitem[Shore et al.(2013a)]{sh13b} Shore, S.~N., Schwarz, G. J., Alton, K., et al.\ 2013a, The Astronomer's Telegram, \# 5409
\bibitem[Shore et al.(2013b)]{sh13} Shore, S.~N., Skoda, P., Korcakova, D., et al.\ 2013b, The Astronomer's Telegram, \# 5312 %% spectroscopy of Del 2013
\bibitem[Shore(2013)]{sh13c} Shore, S.~N.\ 2013, The Astronomer's Telegram, \# 5410
\bibitem[Sitarek \& Bednarek(2012)]{sb12} Sitarek, J., \& Bednarek, W.\ 2012, \prd, 86, 063011 
\bibitem[Stanek et al.(2013)]{st13} Stanek, K.~Z., Shappee, B. J., Kochanek, C. S., et al.\ 2013, The Astronomer's Telegram, \# 5186 
\bibitem[Starrfield et al.(2012)]{st12} Starrfield, S., Iliadis, C., Timmes, F.~X., et al.\ 2012, Bulletin of the Astronomical Society of India, 40, 419 %%(...Hix, W. R., Arnett, W. D., Meakin, C., Sparks, W. M.)
\bibitem[Tatischeff \& Hernanz(2007)]{th07} Tatischeff, V., \& Hernanz, M.\ 2007, \apjl, 663, L101 %% particle acceleration in symbiotic novae
\bibitem[Woudt \& Ribeiro(2014)]{wr14} Woudt, P. A. \& Ribiero, V. A. R. M. (eds) 2014, Stella Novae: Past and Future Decades. ASP Conference Series, Vol. 490 (Astron. Soc. Pacific)
\bibitem[Zanin et al.(2013)]{magic_mars} Zanin, R., Carmona, E., Sitarek, J., et al., 2013, Proc of 33rd ICRC, Rio de Janeiro, Brazil, Id. 773
\end{thebibliography}
\end{document}